\documentclass[12pt]{iopart}
\usepackage{epsf}  
\begin{document}

\title{
System-Size Dependence  of Strangeness
Saturation.
}
\author{
{\sc J. Cleymans$^a$,
B. K\"ampfer$^b$,
P. Steinberg$^a$\footnote{Visiting Fulbright Professor on leave of absence from the Brookhaven National Laboratory, Upton, NY, USA},
S. Wheaton$^a$}
}
\address{
$^a$ Department of Physics, University of Cape Town,
Rondebosch 7701, Cape Town, South Africa\\
$^b$ Institut f\"ur Kern- und Hadronenphysik,
Forschungszentrum Rossendorf, PF 510119, D-01314 Dresden, Germany\\
}

\begin{abstract} 
The final state  in
heavy-ion collisions has a higher
degree of strangeness saturation than the one
produced in collisions between elementary 
particles like $p-p$ or $p-\bar{p}$. 
A systematic analysis of this phenomenon is made for $C-C$, $Si-Si$ and 
$Pb-Pb$ collisions at 
the CERN SPS collider and for $Au-Au$ collisions
 at RHIC and at AGS energies. 
Strangeness saturation is shown to increase smoothly 
with the number of participants at AGS, CERN and RHIC energies.
\end{abstract}
Statistical-thermal models are able 
to fit the multiplicities measured in 
relativistic heavy-ion collisions with remarkable success
\cite{review,stachel,abundances_a,abundances_b}.
A striking feature is that 
 the freeze-out temperatures 
observed in $p-p$, $p-\bar{p}$ and in relativistic heavy-ion collisions
are  similar but the strangeness saturation is very different.  
In this paper we investigate this difference and determine
the thermal parameters as a function of  the
number of participants. We conclude that the 
strangeness saturation 
increases smoothly with the size of the system 
at all energies~\cite{we1,we2,NuXu,Kaneta}. 

The dependence on the
the system size is deduced from
4$\pi$-yields measured in
central $C-C$ and $Si-Si$ collisions~\cite{C_Si}, and 
centrality-binned $Pb-Pb$ collisions~\cite{Sikler,Blume} 
at 158 AGeV at the CERN-SPS. 
For comparison, results from centrality-binned 
mid-rapidity yields from $Au-Au$ collisions at 
$\sqrt{s}_{NN} = 130$ 
GeV~\cite{PHENIX} are also shown. 

\begin{figure}[tbh]
\centerline{
%\vspace{20pt}
\epsfysize=22pc
\epsfxsize=22pc
\epsfbox{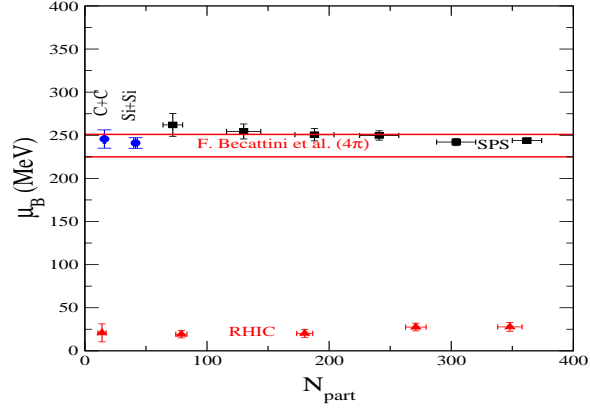}}
\vspace{-45pt}
\caption{
The system-size dependence of the baryon chemical potential, $\mu_B$, 
as extracted from centrality-binned $Pb-Pb$~\cite{Sikler,Blume}, and 
central $C-C$ 
and $Si-Si$ data~\cite{C_Si}. Also shown are the points 
from RHIC~\cite{PHENIX}
}
\label{cleymans_ps3}
\end{figure}

\begin{figure}[tbh]
\centerline{
\epsfysize=22pc
\epsfxsize=22pc
\epsfbox{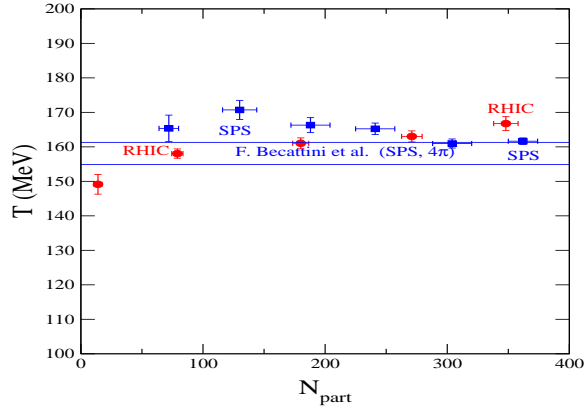}}
\vspace{-45pt}
\caption{
The system-size dependence of the chemical freeze-out temperature, 
as extracted from centrality-binned $Pb-Pb$~\cite{Sikler,Blume}
and from $Au-Au$~\cite{PHENIX}.
}
\label{T_syssize}
\end{figure}

\begin{figure}[tbh]
\centerline{
\vspace{20pt}
\epsfysize=22pc
\epsfxsize=22pc
\epsfbox{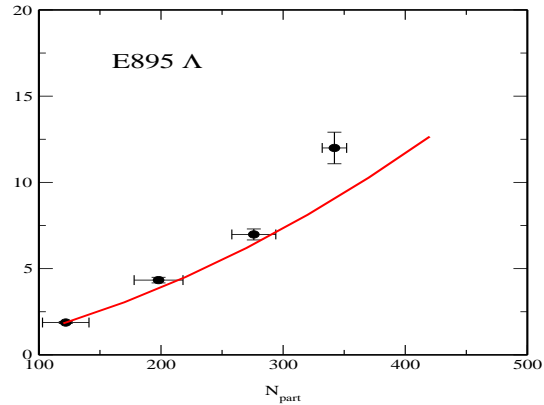}}
\vspace{-45pt}
\caption{The centrality dependence of $\Lambda$ as measured
by the E895 collaboartion~\cite{chung}. The solid line is the
result of a thermal model calculation using a linearly
increasing strangeness saturation factor.}
\label{e895la}
\end{figure}

\begin{figure}[tbh]
\centerline{
\vspace{20pt}
\epsfysize=22pc
\epsfxsize=22pc
\epsfbox{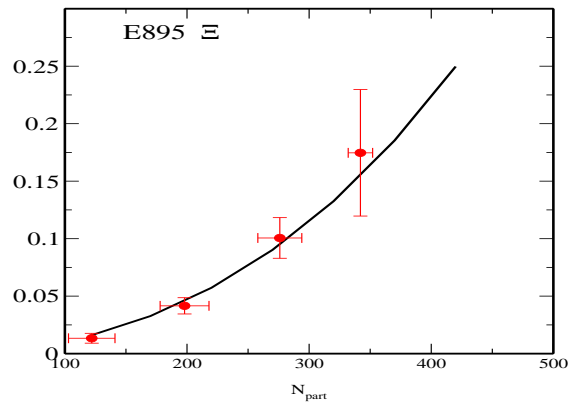}}
\vspace{-45pt}
\caption{The centrality dependence of $\Xi$ as determined by the 
E895 collaboration~\cite{chung}. The solid line has been obtained
using the same parameters is those used in Fig.~4.}
\label{e895ksi}
\end{figure}

The baryon chemical potential, $\mu_B$ is shown in Fig.~\ref{cleymans_ps3}.
It is remarkable that $\mu_B$ shows no dependence at all on 
centrality both at SPS and at RHIC energies. 
The $C-C$ and $Si-Si$ values
are consistent with those obtained in $Pb-Pb$  at the
same energy. For comparison we indicate
in Fig.~\ref{cleymans_ps3} the
results obtained in a comprehensive analysis of all
data at CERN-SPS~\cite{heavy_ions}. 
The freeze-out temperature, $T$, is shown in Fig.~\ref{T_syssize}.
It is again noticeable that $T$ shows almost no dependence on 
centrality both at SPS and at RHIC energies but the 
evidence is less pronounced than for $\mu_B$. 
At RHIC the evidence is compatible with
a smooth increase in the temperature as the centrality increases. 
For comparison we indicate again  the
results of the  comprehensive analysis of reference~\cite{heavy_ions}. 
\vfill\eject
%figure 1%%%%%%%%%%%
%
%figure 2%%%%%%%%%%%
The clearest change in the thermodynamic parameters,
as one changes the size of the system,
is seen in the 
strangeness saturation factor~\cite{Rafelski}, $\gamma_s$, 
which shows a
smooth linear increase 
with  centrality in the $Pb-Pb$ and $Au-Au$ system, except for 
the two most 
central bins in $Pb-Pb$ (see Fig.~\ref{cleymans_ps3}). 
The $C-C$ and 
$Si-Si$ systems lie above the trend suggested by the $Pb-Pb$
 points. 
This clearly indicates that peripheral $Pb-Pb$ collisions are not 
equivalent, with respect to strangeness saturation, to central 
collisions of lighter nuclei with the same 
participant number. 
%
%figure 3%%%%%%%%%%%
\begin{figure}[tbh]
\centerline{
\vspace{20pt}
\epsfysize=22pc
\epsfxsize=22pc
\epsfbox{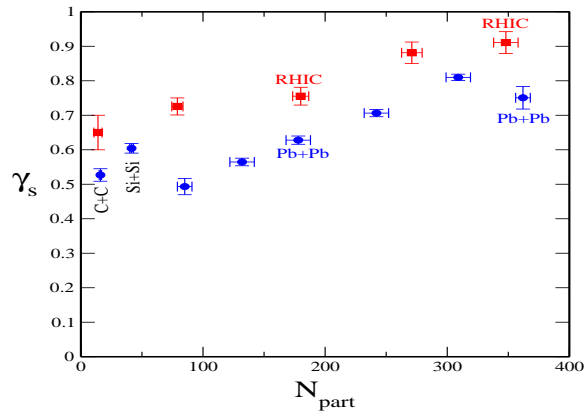}}
\vspace{-40pt}
\caption{
Dependence of the strangeness saturation factor $\gamma_s$ as a function
of the number of participants. The lower (round) points refer to CERN-SPS
at 156 GeV beam energy while the higher (square) points refer to RHIC $Au-Au$.
}
\label{cleymans_ps5}
\end{figure}
At this conference first results were presented by the E895 collaboration
\cite{chung} on $\Xi$ and $\Lambda$ yields obtained in $Au-Au$
collisions at the AGS. These data require both canonical 
corrections and a strangeness saturation which increases
linearly with the  number of participants, which was taken as 
$\gamma_s = 0.32 + 0.0015N_{\mathrm part}$, in rough agreement with
the results from SPS. The fit
shown in Figs.~\ref{e895la}, \ref{e895ksi} uses the 
freeze-out temperature and chemical potential obtained from the 
values as determined from the fit \cite{heavy_ions}. 

In conclusion, the strangeness saturation factor, $\gamma_s$,
increases with participant number in the $Pb-Pb$ system at the 
CERN SPS as well as the $Au-Au$ system at RHIC. 
Central collisions of $C-C$ 
and $Si-Si$ at SPS energies deviate, with respect to strangeness 
saturation, from peripheral $Pb-Pb$ collisions. 
%
%
%figure 4%%%%%%%%%%%
%
%\newpage
%
%
%
%figure 7%%%%%%%%%%%
%
%
%
%figure 8%%%%%%%%%%%
\section*{Acknowledgments} 
We acknowledge  useful discussions with C. H\"ohne, 
M. Ga\'zdzicki and P. Chung. Two of us (J.C. and S.W.) acknowledge
 the financial assistance of 
the National Research Foundation (NRF) of South Africa and of the
URC of the University of Cape Town.    

\section*{References}
%===================================
 
%-----------------------------------------------------------
%

\begin{thebibliography}{999}
%===================================
\bibitem{review} For a general review see e.g. 
K. Redlich, J. Cleymans, H. Oeschler and A. Tounsi, 
Acta Physica Polonica B33 (2002) 1609.
%
\bibitem{stachel} P. Braun-Munzinger, K. Redlich and J. Stachel, 
to appear in "Quark Gluon Plasma 3", eds. R.C. Hwa and Xin-Nian Wang, 
World Scientific Publishing, nucl-th/0304013.
%
\bibitem{abundances_a} 
P. Braun-Munzinger et al., Phys. Lett. 
B 344 (1995) 43, 
B 365 (1996) 1,
B 465 (1999) 15,
B 518 (2001) 415.
\bibitem{abundances_b}
J. Cleymans, K. Redlich, Phys. Rev. Lett. 81 (1998) 5284;\\
J. Sollfrank, J. Phys. G: Nucl. Part. Phys. 23 (1997) 1903.
%
\bibitem{we1} J. Cleymans, B. K\"ampfer, S. Wheaton, 
Phys. Rev. C 65 (2002) 027901.
\bibitem{we2} J. Cleymans, B. K\"ampfer, P. Steinberg, S. Wheaton, 
hep-ph/0304269 and hep-ph/0212335.
\bibitem{NuXu} N. Xu et al. (STAR collaboration), private communication.
\bibitem{Kaneta} M. Kaneta, private communication (2003).
\bibitem{C_Si} C. H\"ohne (NA49 collaboration), nucl-ex/0209018. 
\bibitem{Sikler} F. Sikler (NA49 collaboration), Nucl. Phys. A 661
(1999) 45c.
\bibitem{Blume} V. Friese et al. (NA49 collaboration), 
Nucl. Phys. A 698 (2002) 487c.
\bibitem{PHENIX} K. Adcox et al. (PHENIX collaboration), Phys. Rev. Lett. 88 (2002) 242301. 
\bibitem{heavy_ions}F. Becattini, J. Cleymans, A. Ker\"anen, E. Suhonen, K. Redlich,
Phys. Rev. C 64 (2001) 024901.
%
\bibitem{Rafelski} J. Letessier, J. Rafelski and A. Tounsi, Phys. Rev. C 50 
(1994) 406;\\
C. Slotta, J. Sollfrank and U. Heinz, in {\sl Proceedings
of Strangeness in Hadronic Matter}, Tucson, edited 
by J. Rafelski, AIP Conf. Proc. No. 340 (AIP, Woodbury, 1995), p. 462.
\bibitem{Hoehne_private} C. H\"ohne, private communication (2002).
%
\bibitem{chung}P. Chung et al., E895 collaboration, nucl-ex/0302021.
\end{thebibliography}
\end{document}